# Single-photon decision maker

Makoto Naruse[1], Martin Berthel[2,3], Aurélien Drezet[2,3], Serge Huant[2,3], Masashi Aono[4,5], Hirokazu Hori[6] & Song-Ju Kim[7]

1 Photonic Network Research Institute, National Institute of Information and Communications Technology, 4-2-1 Nukui-kita, Koganei, Tokyo 184-8795, Japan

2 Université Grenoble Alpes, Inst. NEEL, F-38000 Grenoble, France

3 CNRS, Inst. NEEL, F-38042 Grenoble, France

4 Earth-Life Science Institute, Tokyo Institute of Technology, 2-12-1 Ookayama, Meguru-ku, Tokyo 152-8550, Japan

5 PRESTO, Japan Science and Technology Agency, 4-1-8 Honcho, Kawaguchi-shi, Saitama 332-0012, Japan

6 Interdisciplinary Graduate School of Medicine and Engineering, University of Yamanashi, Takeda, Kofu, Yamanashi 400-8511, Japan

7 WPI Center for Materials Nanoarchitectonics, National Institute for Materials Science, 1-1 Namiki, Tsukuba, Ibaraki 305-0044, Japan

**Abstract:** Decision making is critical in our daily lives and for society in general and is finding evermore practical applications in information and communication technologies. Herein, we demonstrate experimentally that single photons can be used to make decisions in uncertain, dynamically changing environments. Using a nitrogen vacancy in a nanodiamond as a single-photon source, we demonstrate the decision-making capability by solving the multi-armed bandit problem. This capability is directly and immediately associated with single-photon detection in the proposed architecture, leading to adequate and adaptive autonomous decision making. This study makes it possible to create systems that benefit from the quantum nature of light to perform practical and vital intelligent functions.

The need to make accurate decisions in uncertain, dynamically changing environments appears in all aspects of life. Which stock should I buy? Which option should I take in the next move in a board or electronic game? These seemingly casual scenes are highly related to a Monte Carlo tree search[1,2], which is a heuristic search algorithm for decision making. Consider, e.g., an information network infrastructure. Ever-increasing demands in mobile communications outstrip the available radio frequencies; consequently, dynamic and adequate frequency assignment is critical. This scenario constitutes a decision-making problem in an uncertain environment[3,4]. As another example, consider maximizing the revenue for an e-commerce website on the Internet; this requires presenting adequate content and advertisement in a limited screen space of typical displays and is also a corollary of decision-making problems[5].

Several computing algorithms such as ε-greedy[6], softmax[6,7], upper confidence bounds[8], and tug-of-war (TOW)[4,9,10] have been proposed in the literature to solve these decision-making problems. All these algorithms have been designed essentially using *probabilistic* mechanisms to resolve the "*exploration–exploitation dilemma*" tradeoff in decision making. The present study aims to *physically* implement decision making using the intrinsic *quantum* attributes of *single photons*.

Actually, the softmax rule, which is based on the notion of the Boltzmann factor, has previously been discussed as the best-fitting algorithm for human decision making[7]; here, we may see a correspondence between the physical mechanisms of the brain and decision making. Indeed, investigating the relation between physics and decision making with an aim to better understand the origin of intellectual abilities of natural organisms would be highly stimulating. The TOW algorithm[9] was inspired by the spatiotemporal dynamics of the slime mold *Physarum*[11]. More specifically, the idea is originally based on observing living organisms in nature and their interactions with various environments. This fact implies that it is possible to engineer *artificially constructed decision-making machines*, which would contribute significantly to resolving decision-making problems in practical information and communications technology (ICT). In fact, Kim *et al.* (2013) proposed a theory of a TOW-based decision maker that uses nanoscale optical excitation transfer among quantum dots mediated by near-field interactions[12]. They followed this study by an experimental verification based on colloidal quantum dots of different sizes[13]; this paved the way to the implementation of a solid-state, ultrasmall decision maker. Moreover, the TOW-based physical method outperforms other algorithms[12].

However, many important unresolved problems remain before we can claim completely autonomous, physical decision-making machines. One of the most critical concerns is that the probabilistic mechanism, which is an indispensable attribute in solving decision-making problems, is yet to be realized experimentally. For instance, in the experimental demonstration reported in Naruse *et al.* (2014)[13], the probabilistic decision-making step was implemented by an electrical host controller and the probability was determined by observing optical energy transfer in ensembles of quantum dots.

In this study, we report decision making based on *single photons*. One of the most significant attributes of single photons is literally its particle and yet probabilistic nature. For example, consider a single photon that impinges on a beam splitter. The probability of observing the single photon in one of the two output channels is 50%. However, in detecting individual events, when a single photon is detected in one channel, the other channel does *not* detect it. We see here a TOW-type mechanism in the sense that an input photon is pulled by one channel and by the other, which will be discussed in detail in this study. Observing a single photon can be *directly* associated with decision making. In other words, without requiring emulation by electrical computing, the decisive step is directly implemented by the intrinsic attributes of a single photon and is fundamentally unachievable using classical light.

In addition, whereas the use of single photons in ICT has thus far been limited to the context of quantum key distributions[14] and quantum computing[15], including quantum simulations[16], this study paves the way to realizing the benefits offered by the quantum nature of light in practical and important decision-making applications in ICT. Meanwhile, theoretical quantum-mechanical approaches to machine learning are emerging[17,18], and we hope to contribute to the field by presenting an architecture to implement single-photon-based decision making and its experimental demonstration. In the sections below, we describe the architecture, principles, experimental systems, demonstrations, and analysis of the proposed single-photon decision maker.

**Results**

**Architecture and principle**

For the simplest example of the decision-making problem that still preserves the essence of the problem, consider a case in which a player must select either **slot machine L** or **slot machine R**, with the goal being to maximize reward. The reward probability of **slot machines L** and **R** is given by $P_L$ and $P_R$, respectively. Note that $P_L$ and $P_R$ may change over time. To attain the goal, the player may test to know which is better; however, too much testing may result in excessive loss. This exemplifies the tradeoff between "exploration" and "exploitation," which is referred to as the "*exploration–exploitation dilemma*"[7]; such a problem is described by the *multi-armed bandit problem*[19,20], which is the foundation of many important applications in ICT[3,4,5]. In this study, we assume that the rewards dispensed from **slot machine L** and from **slot machine R** are the same so that making accurate decisions means finding a machine with a higher reward probability as quickly as possible.

As mentioned previously, the idea of the TOW-based approach[9,12] came from observing slime molds, which maintain a constant intracellular resource volume while collecting environmental information by concurrently expanding and shrinking their branches. The conservation law entails a nonlocal correlation among the branches, i.e., the volume increment in one branch is immediately compensated by volume decrements in other branches. The TOW is a metaphor to represent such a nonlocal correlation, which enhances the decision-making performance[12].

This mechanism matches well the intrinsic attributes of a single photon. Prepare a polarizing beam splitter (PBS) as shown in Fig. 1(i), in which vertically polarized light is directed to Channel 0 (Ch.0) and horizontally polarized light is directed to Channel 1 (Ch.1).

(i) When the linear polarization of the input single photon is oriented at 45° with respect to the horizontal, the probability of the photon to be detected in Ch.0 or Ch.1 is 0.5. However, the probability of detection in either Ch.0 or Ch.1 is unity. In the drawing in Fig. 1(i), the photon is detected in Ch.1.

(ii) When the polarization is nearly horizontal, as shown in Fig. 1(ii), detection in Ch.1 will mostly result.

(iii) When the polarization is nearly vertical, as shown in Fig. 1(iii), detection in Ch.0 will mostly result.

The probabilistic attribute of a single photon is obviously represented in case (i), but also in cases (ii) and (iii), where the possibility to be detected in the opposite channel is not perfectly zero because the polarizations are not perfectly horizontal or vertical.

We associate the detection of a photon in Ch.0 or Ch.1 *immediately* with the a decision to select **slot machine R** or **slot machine L**, respectively, as shown in Fig. 2. This is a remarkable aspect of the single-photon decision maker; it exploits the *quantum* attributes of photons. If photon observation was based on *classical* light, e.g., observing the light intensity in Ch.0 and Ch.1, we must implement one additional step to make a "*decision*." This physical fundamental difference between classical and quantum light is decisive in realizing physical decision makers.

In our architecture, the *TOW* mechanism is implemented by the notion of a *polarization adjuster* (PA), which is shown schematically in Fig. 2. This architecture provides the characteristics as follows:

1. A PA value of zero indicates a polarization at 45° with respect to the horizontal, which corresponds to situation (i) discussed above.

2. The decision to select the designated slot machine is immediately made by observing a single photon in Ch.0 or Ch.1.

3. If a reward is successfully dispensed from the selected slot machine, the PA is "moved" in the direction of the selected machine, i.e., if **slot machine R** is selected and a reward is obtained, the PA is moved so that the input polarization is more *horizontally* polarized. Moreover, if no reward is dispensed from **slot machine R**, the PA is moved in the direction of the unselected machine (i.e., the input polarization is more *vertically* polarized in this case).

By iterating steps 2 and 3, the PA guides us to a decision in which we select the *correct solution*; this means that the slot machine with a higher reward probability will highly likely be selected. An experimental realization of this scheme is discussed in the next section.

**Measurements**

A single photon emitted from a single individual nitrogen-vacancy (NV) colour centre[21], which has broadband emission in the visible range (650–700 nm)[22], passes through a polarizer and then through a zero-order half-wave plate and impinges on the PBS. The two detectors used are avalanche photodiodes (APDs). The detected signal is sent to a time-correlated single-photon-counting (TCSPC) system where individual events are directly captured. The details of the optical system used in the experiment and single-photon emission from the NV centre are described in the Methods section and in the Supplementary Information.

The sequence of single photons detected in Ch.0 and Ch.1 is shown in Fig. 3. The vertical red and blue bars represent single-photon detection events, and the horizontal axis represents time. Here, we introduce the notion of "play" or "cycle." A single play of the selected slot machine is referred to as a play, and the time to complete a single play is defined as a cycle. Multiple single photons may be emitted from the single-photon source during a cycle, as shown in Fig. 3. The *decision* is made by the *first single-photon detection* event of a given single play or cycle. In the case shown in Fig. 3, a single photon is first detected in Ch.1 in the first play, leading to the decision of selecting **slot machine R**. In the second and third plays, a single photon is first detected in Ch.0, leading to the decision of selecting **slot machine L**. In parallel to the experiment, the slot machines were emulated by a host controller. More specifically, the reward probability $P_L$ and $P_R$ were given as threshold values. If a random number between 0 and 1 generated by the host controller is less than the reward probability of the selected slot machine, the reward is dispensed. In the experimental procedure (see [E1]–[E6] below), this process of slot-machine playing and the subsequent reward-receipt process corresponds to [E2] and [E3], respectively.

On the basis of the PA value, the linear polarization is made more vertical or horizontal by rotating the half-wave plate using a rotary positioner. In Fig. 4, the red crosses and blue circles indicate two seconds worth of photon counts detected in Ch.0 and Ch.1, respectively, as a function of the orientation of the half-wave plate. Note that the orientation angle shown in Fig. 4 does not indicate linear polarization with respect to the horizontal direction but refers to the absolute value defined in the rotary positioner used in the experiment. Because the sensitivities of the APDs in Ch.0 and Ch.1 are not precisely equal and owing to possible misalignment in the optical setup, Ch.0 counts more photons than Ch.1 at an angle that yields the maximum counts. Three snapshots of the observed sequence of single photons acquired over 10 μs with the half-wave plate selecting polarization angles of 20°, 35°, and 60° are shown at the top of Fig. 4. The 20° and 60° polarization angles correspond to the input photon being nearly completely horizontally polarized and nearly completely vertically polarized, respectively, whereas the 35° polarization angle corresponds to approximately the 50:50 situation. In addition, the extinction ratio of the polarizer is $10^5$ and that of PBS is $10^3$ (the product information is shown in the Methods section). We consider that intrinsic optical properties of various optical components in the experimental setup do not lead to any significant asymmetry.

To realize the PA mechanism in our architecture, we quantify the TOW mechanism as follows: Let the initial PA value be $PA_0$. If, in cycle *t*, the selected machine is **slot machine L** and yields a reward *or* the selected machine is **slot machine R** and loses the play, the PA value is updated at cycle *t* + 1 based on

$$PA(t+1) = -\Delta + PA_0 + \alpha \left[ PA(t) - PA_0 \right], \tag{1}$$

where $\alpha$ is referred to as the "forgetting parameter"[12] and $\Delta$ is the constant PA increment (in this experiment, $\Delta$ = 1 and $\alpha$ = 0.99). Equation (1) indicates that a more *vertical* input polarization results from smaller PA values. If, in cycle *t*, the selected machine is **slot machine R** and yields a reward *or* the selected machine is **slot machine L** and loses the play, the PA value is updated based on

$$PA(t+1) = +\Delta + PA_0 + \alpha \left[ PA(t) - PA_0 \right], \tag{2}$$

which means that a more *horizontal* input polarization results from larger PA values.

On the basis of the actual photon-counting statistics obtained from Fig. 4, the orientation of the linear polarization for the single-photon input at cycle *t* is

$$P(t) = \text{Pos}\left( \lceil PA(t) \rceil \right), \tag{3}$$

where $\lceil \; \rceil$ means the floor function, which truncates the decimal part. The function Pos(*n*) specifies the orientation of the half-wave plate when the truncated PA value is *n* based on the given experimental condition. For example, referring to Fig. 4, the PA value of 0 corresponds to approximately 35°, i.e., $\text{Pos}(0) \approx 35$.

Thus, for the above described experimental system, the decision-making procedure is summarized as follows:

[E1]   Photon detection by Ch.0 (directly associated with the decision to select **slot machine L**) or Ch.1 (directly associated with the decision to select **slot machine R**) through APDs and a TCSPC system.

[E2]   Play the selected slot machine.

[E3]   Reward dispensed or not dispensed.

[E4]   Update the PA value on the basis of Eq. (1) or (2).

[E5]   Determine the orientation of the half-wave plate using Eq. (3).

[E6]   Control the rotary positioner, and then, return to step E1.

The details of the experimental apparatus and the PA are given in the Methods section.

**Discussion**

We now implement two control mechanisms referred to hereafter as **Control 1** and **Control 2**. For **Control 1**, the PA values are associated with $P(t)$, which orients the polarization between 20° and 60°, where the ratio of photon counts detected in Ch.0 to those detected in Ch.1 is maximized and minimized, respectively (see Fig. 4). In **Control 2**, however, the PA values are correlated with $P(t)$, which orients the polarization between 20° and 40°, where the maximum photon counts in Ch.0 and Ch.1 are comparable with each other.

Let the initial reward probability of the slot machines be given by $P_L = 0.8$ and $P_R = 0.2$, which means that selecting **slot machine L** is the correct decision. The initial PA value ($PA_0$) is set to zero. To verify the ability to adapt to environmental changes, the reward probability is inverted every 150 cycles in the experiment. The decision maker played 600 consecutive plays, and these 600 plays were repeated ten times. We determine the *correct selection rate* by calculating the number of correct decisions or by taking the higher-reward probability machine and dividing by the number of repeat cycles. The solid curve in Fig. 5a gives the correct selection rate based on the **Control 1** policy, which gradually increases with time. Because of the sudden swapping of the reward probability, the correct selection rate drops every 150 plays but quickly recovers. Such rapid and accurate adaptability demonstrates that the single-photon decision maker can solve the multi-armed bandit problem. Meanwhile, the solid curve in Fig. 5b shows the time evolution of the PA value, which agrees well with the given environments. In other words, the PA value *decreases* from cycle 0 to 150 because the selection of **slot machine L** is the correct decision, which is more likely to be the decision taken using large, negative PA values. Conversely, the PA value *increases* from cycle 151 to 300 because the selection of **slot** **machine R** is now the correct decision, and this decision is more likely to be taken using large, positive PA values.

The correct selection rate reaches unity rather quickly in the first region (from cycle 1 to 150), but for the subsequent regions (e.g., cycle 151 to 300), the rate takes approximately 50 cycles to recover. This is consistent with the evolution of the PA value, which was initially zero ($PA_0 = 0$) in the first region and decreases to approximately −45 at cycle 150. Adapting to the changing reward probability, the PA value increases after cycle 151 to positive values and requires a few cycles to reach the goal. This phenomenon explains the different number of elapsed cycles required to reach the correct selection rate of unity. After a sudden change in the environment, the slope of the correct selection rate depends on $\alpha$ [the forgetting parameter; see Eqs. (1) and (2)]. For smaller $\alpha$, the adaptation is quicker. However, smaller $\alpha$ also prevents the correct selection rate from reaching unity. In other words, quick adaption to environmental changes is a tradeoff with maximizing the correct selection rate. The physical time required in each cycle is discussed at the end of the paper.

The dotted curve in Fig. 5a shows the correct selection rate based on the **Control 2** policy, which performs poorly compared with **Control 1**. This is also clearly observed by evaluating the average of the correct selection rate, which is shown in the inset of Fig. 5a. In particular, the **Control 2** policy is inferior to the **Control 1** policy during cycles 1 to 150 and 301 to 450, where selecting **slot machine L** is the correct decision. During these plays, the PA value should decrease, as indeed it does (see Fig. 5b). However, the smallest PA value for the **Control 2** policy corresponds to a 40° orientation of the half-wave plate, where the difference between photon counts in Ch.0 (in this case, corresponding to correct decisions) and Ch.1 (incorrect decisions) is less evident. This is another remarkable achievement of the experiment in the sense that the inherent physical imbalance of the system, which is because of the quantum nature of light (see Fig. 4), is manifested by the clear difference in terms of decision-making performance, i.e., quantum attributes lead clearly to the decision-making ability.

Considering the **Control 1** policy, the dotted curve in Fig. 6a shows the evolution of the correct selection rate with initial reward probability $P_L = 0.6$ and $P_R = 0.4$, which are also inverted every 150 plays. Because the difference between the reward probability is less than that in the former case, as shown by the solid curve in Fig. 6a, making accurate decisions becomes more difficult, which is manifested by the fact that the dashed curve remains mostly below the solid curve in Fig. 6a. However, remarkably, the correct selection rate again increases with time and such adaptive behaviour is also observed in such a difficult environment. Corresponding to the increased difficulty of the situation, the fluctuation of the PA values, shown by the dotted curve in Fig. 6b, is less evident than in the former problem (cf. Fig. 5b). At the same time, we can also argue that smaller fluctuations are advantageous for adapting to environmental changes. As observed by the behaviour immediately after the reward probability is inverted at 150, 300, and 450 cycles, the correct selection rate recovers more quickly, and in fact, the resulting correct selection rate outperforms the former case.

Finally, we make a few remarks about the present study including some practical notes. First, we note how multiple photons affect the system. If Ch.0 and Ch.1 simultaneously detect photons, a decision *cannot* be made. In other words, a multi-photon measurement leads to an error from the overall decision-making system. Whereas the second-order photon-intensity correlation $g^{(2)}(0)$ is greater than zero (see Supplementary Information for details), no error occurred in the experimental demonstration. This aspect of the demonstration is another signature of decision making based on single photons.

The second remark concerns the latency of the experimental system. As described by step [E1], the timing of the arrival of single photons is determined by the APDs followed by a TCSPC system. Although our system can capture a single-photon series with 10 million counts/s *without loss*, the "latency" in data acquisition cannot be defined or measured because the TSCPC system in use is *not* a real-time system. The latency varies as a function of various internal factors. Steps [E2]–[E5] are computed at the host controller, which requires approximately 1 ms in total. More precisely, the host controller in use is also *not* a real-time system (see Methods section for the specifications). However, the processing time is experimentally measured and is approximately 1 ms. In step [E6], the latency of controlling the rotary positioner to configure the angle of the half-wave plate depends on the traveling angle. For the **Control 1** policy, for example, the maximum travel angle in one cycle is 24°(see Methods section for details), which takes approximately 3.5 seconds. When the traveling angle is small (for instance, 1°), the latency is approximately 0.7 seconds. Overall, the total latency for a single cycle is of the second order, with the primal latency being attributed to the mechanical polarization control. We consider that the speed with which the polarization is controlled can be significantly improved by introducing an electro-optic phase modulator. If the data acquisition, post processing, and polarization control processes were sufficiently fast, then the bottleneck would become the rate of the single-photon source, which is currently approximately 5000 photons/second (see Supplementary Information).

The third remark concerns the single-photon acquisition and the controlling system. As described above, the duration of a single cycle depends on the status of the system. Because our primary interest is to demonstrate the principle of single-photon-based decision making, the issue of *hard real-time* operation[23], in which all deadlines are strictly smaller than a specified value, is *not* a critical concern in the present study. However, hard real-time operation in photonic systems is important for the future development of this single-photon-based experimental system.

Finally, we discuss the scalability of the system. If the number of the slot machines becomes $N$, the setup would require $N-1$ polarization controllers and PBSs and $N$ photodetectors. In these scenarios, adapting integrated photonics technologies, previously used in some quantum systems[24,25], is an interesting topic for future study. Another exciting direction of research is the use of single-photon nanophotonic devices[26] based on optical-near-field-mediated energy transfer[13]. In addition, the use of novel architectures to reduce hardware complexity merits future study.

To summarize, we propose herein an architecture for a *single-photon decision maker* and experimentally demonstrate accurate and adaptive decision making using the NV centre in a nanodiamond as a single-photon source. Because of the quantum nature of light, single-photon detection is immediately and directly associated with decision making, which is a decisive step to achieve an autonomous intelligent machine based on a purely physical mechanism. In addition, this work paves the way to exploit the characteristics of the quantum nature of light in practical and vital intelligent roles in ICT, which contrasts with the current discussions of single photons in the literature that focus only on the context of quantum key distributions[14] and quantum computing[15]. In parallel with our research, implementing intelligent functions based on the attributes of light has been intensely investigated[27–30]. The fusion of these emerging technologies constitutes an exciting topic for future research.

**Methods**

**Optical system**

The single-photon beam was supplied by a single NV centre in a surface-purified 80-nm nanodiamond[31]. The NV centre was selected by fluorescence imaging and second-order time-intensity correlation measurements[32,33] (see Supplementary Information). This approach was used to guarantee single-photon emission. The excitation laser was operated at the wavelength of 532 nm (green diode-pumped solid-state laser by CNI Lasers). The detectors consisted of APDs (Excelitas Technologies SPCM-AQRH-16-FC) with a peak photon detection efficiency greater than 70% at 700 nm and a background count below 25 counts per second. The TCSPC system was a PicoQuant PicoHarp 300 module with a time-tagged time-resolved mode that enables individual count events to be recorded. The half-wave plate was mounted on a Thorlabs motorized rotary positioner PRM1Z8 driven via a dc servomotor that provides 1 arc-second resolution. We used standard optical components adapted to our spectral range [PBS (CM1-PBS252), half-wave plate (WPH10M-694), and polarizer (LPVISB050-MP2) from Thorlabs; long-pass filter (BLP02-561R-25; cutoff wavelength: 561 nm) from Semrock; and an immersion objective (numerical aperture: 1.4) from Nikon (CFI Pal Apo VC 100X Oil)]. The host controller is Hewlett Packard's HP Z400 with an Intel Xeon W3520 CPU (2.66 GHz), OS Windows 7 professional 64 bits, and 4.00 GB RAM. A LabVIEW-based program was used to control the experimental system, as described in steps [E1]–[E6] in the main text.

**Polarization Adjuster**

For the decision-making experiments, in both cases based on the **Control 1** and **Control 2** policies, the truncated integer values of the PA value, $\lceil PA(t) \rceil$, were assumed to be elements of the set −3, −2, −1, 0, 1, 2, 3. When $PA(t) \geq 4$ or $PA(t) \leq -4$, the truncated PA value is defined by $\lceil PA(t) \rceil = -3$ and $\lceil PA(t) \rceil = 3$, respectively. Note that the PA value itself, $PA(t)$, can take any real number according to the dynamics described by Eqs (1) and (2).

The orientation of the linear polarization based on the truncated integer values of the PA value, as formulated by Eq. (3), is given by

1. **Control 1 policy**: $\text{Pos}(-3) = 60$, $\text{Pos}(-2) = 59$, $\text{Pos}(-1) = 58$, $\text{Pos}(0) = 34$, $\text{Pos}(1) = 21$, $\text{Pos}(2) = 20$, $\text{Pos}(3) = 19$.

2. **Control 2 policy**: $\text{Pos}(-3) = 42$, $\text{Pos}(-2) = 41$, $\text{Pos}(-1) = 40$, $\text{Pos}(0) = 34$, $\text{Pos}(1) = 21$, $\text{Pos}(2) = 20$, $\text{Pos}(3) = 19$.

The function $\text{Pos}(\lceil PA(t) \rceil)$ can be given in other ways, but the resulting decision-making performance does not depend significantly on these differences. An additional experimental result is described in the Supplementary Information.

**Acknowledgements**

This work was supported in part by the Core-to-Core Program, A. Advanced Research Networks from the Japan Society for the Promotion of Science, and in part by Agence Nationale de la Recherche, France, through the SINPHONIE project (Grant No. ANR-12-NANO-0019).

**Author contributions**

M.N., S.H. & S.-J.K. directed the project; M.N. and S.-J.K. designed system architecture; M.N., M.B., A.D. & S.H designed and implemented optical systems; M.B. & M.N. conducted optical experiments; M.N., M.A. & S.-J.K. analyzed data; H.H. discussed physical principles; and M.N., M.B., A.D. & S.H. wrote the paper.

**Competing financial interests**

The authors declare no competing financial interests.

**Author Information**

Correspondence should be addressed to M.N. (naruse@nict.go.jp) or S.-J.K. (KIM.Songju@nims.go.jp).

**Figure captions**

**Figure 1 | Principle of single-photon decision maker: single photon and polarization.** (i) Light polarized linearly with respect to the horizontal axis is incident on a polarizing beam splitter (PBS). The single photon travels down to one of the output channels (Ch.0 or Ch.1) with a probability of 0.5 for each channel. However, the probability of a single photon to travel down one of the channels is unity; thus, the selected channel is immediately associated with a decision. (ii) Nearly horizontally polarized light is highly likely to be detected in Ch.1, leading to the decision of selecting **slot machine R**. (iii) Nearly vertically polarized light is highly likely to be detected in Ch.0, leading to the decision of selecting **slot machine L**.

**Figure 2 | System architecture for single-photon decision maker and schematic of experimental setup.** The excitation laser excites a nitrogen vacancy in a diamond crystal, and the emitted single photon passes through an immersion objective, high-pass filter (F), polarizer, half-wave plate, and PBS and is detected either in Ch.0 or Ch.1. The photon detection is immediately associated with the decision of which machine to play. On the basis of the betting results, the polarization adjuster (PA) controls the orientation of the half-wave plate using a rotary positioner.

**Figure 3 | Correspondence between a series of single photons and decision making in the proposed architecture.**

**Figure 4 | Polarization-dependent photon counts.** Total photon counts over 2 s detected in Ch.0 and Ch.1 as a function of the orientation of the half-wave plate. Upper panels show typical sequences of single photons detected over 10 $\mu$s.

**Figure 5 | Demonstration of single-photon decision maker (1).** (a) The correct selection rate, or the rate of making an accurate decision (i.e., selecting the slot machine with higher reward probability) as a function of time. The correct selection rate increases with time even after the sudden inversion of the reward probability that is induced in the system every 150 cycles. (b) Evolution of the associated PA values.

**Figure 6 | Demonstration of single-photon decision maker (2).** (a) Evolution of the correct selection rate for two different problems. Making an accurate decision with a reward probability of $P_L = 0.6$ and $P_R = 0.4$ is more difficult when compared with $P_L = 0.8$ and $P_R = 0.2$ because the difference in the probability is less. However, autonomous decision making is still achieved. (b) Evolution of the associated PA values.

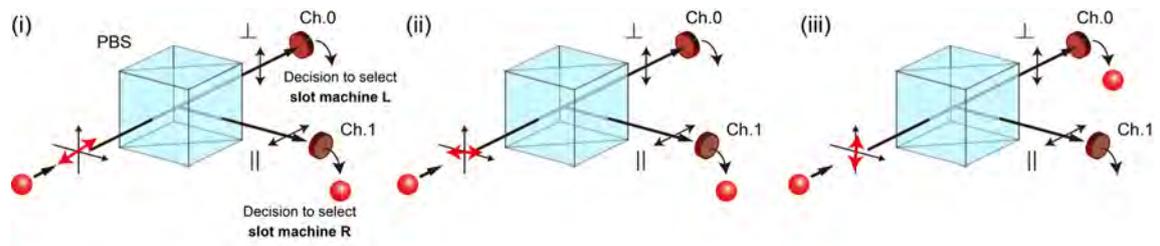

Figure-1 (Naruse)

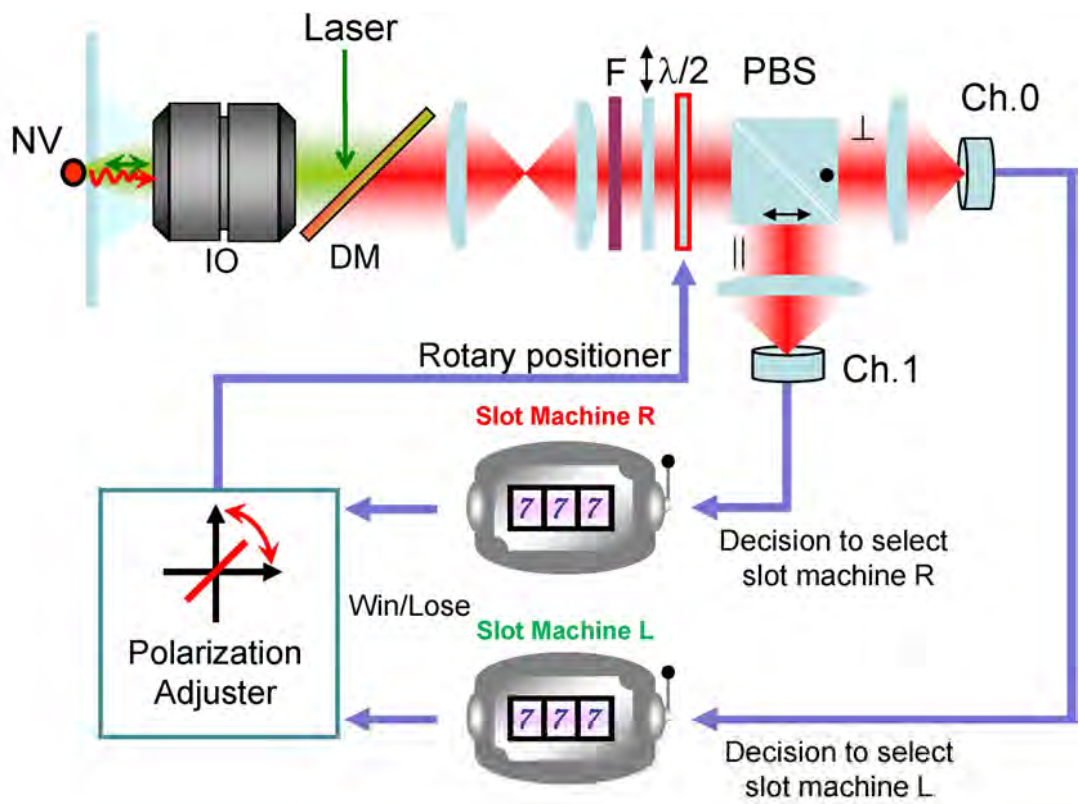

Figure-2 (Naruse)

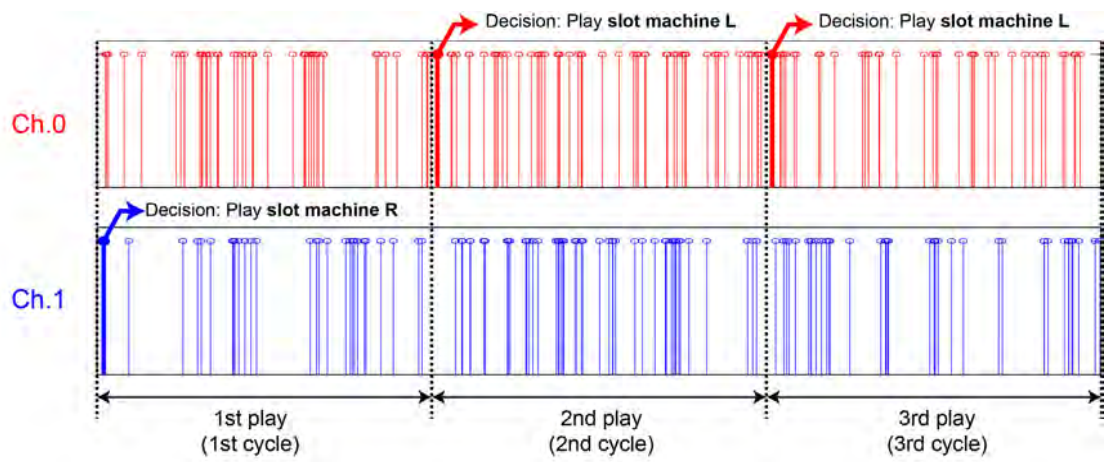

Figure-3 (Naruse)

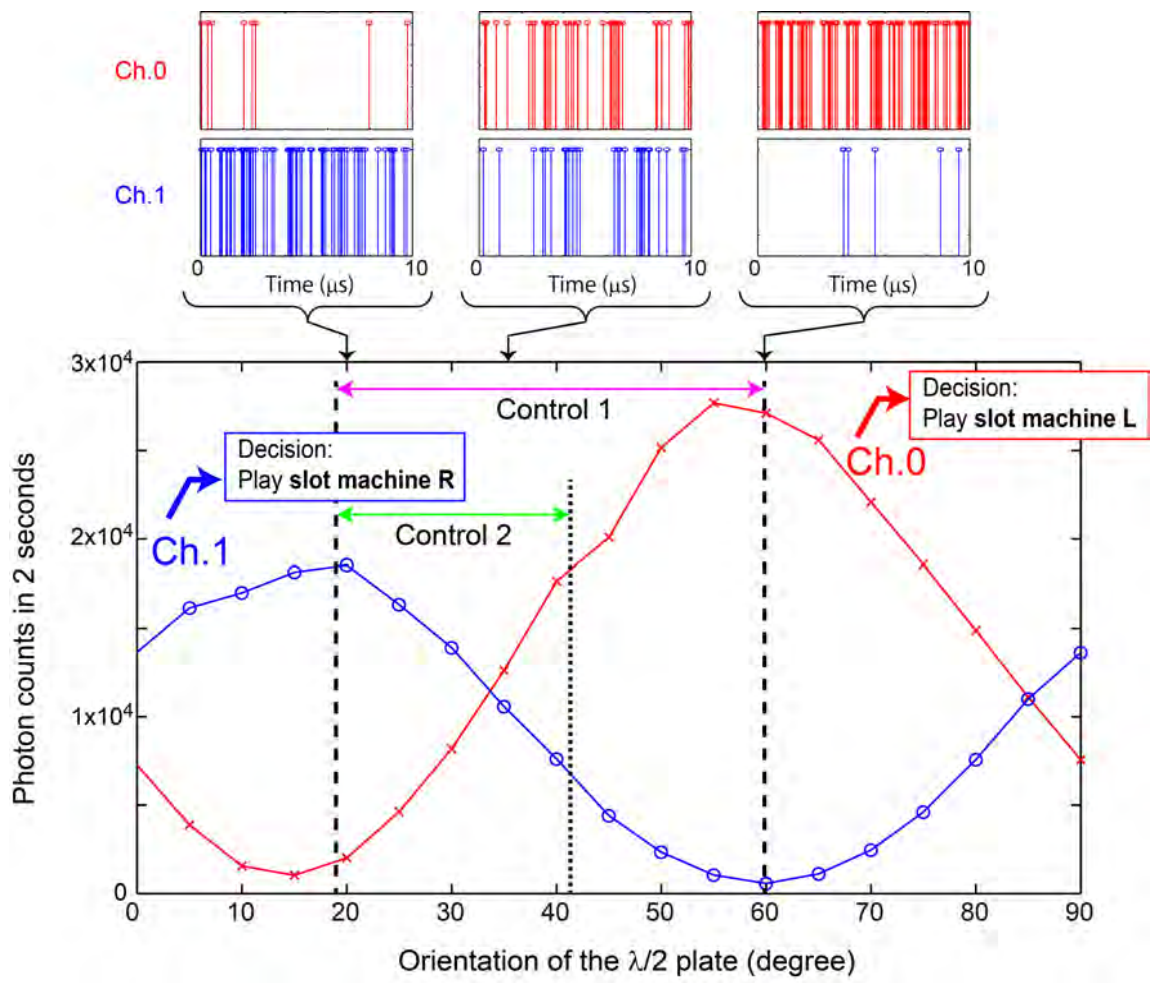

Figure-4 (Naruse)

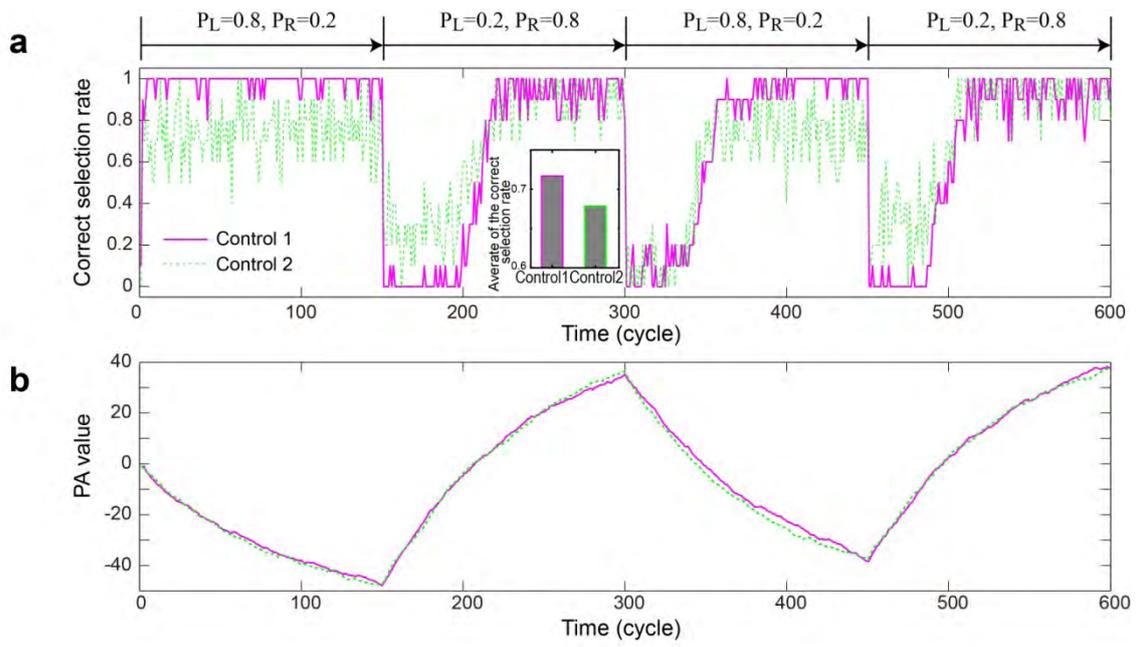

Figure-5 (Naruse)

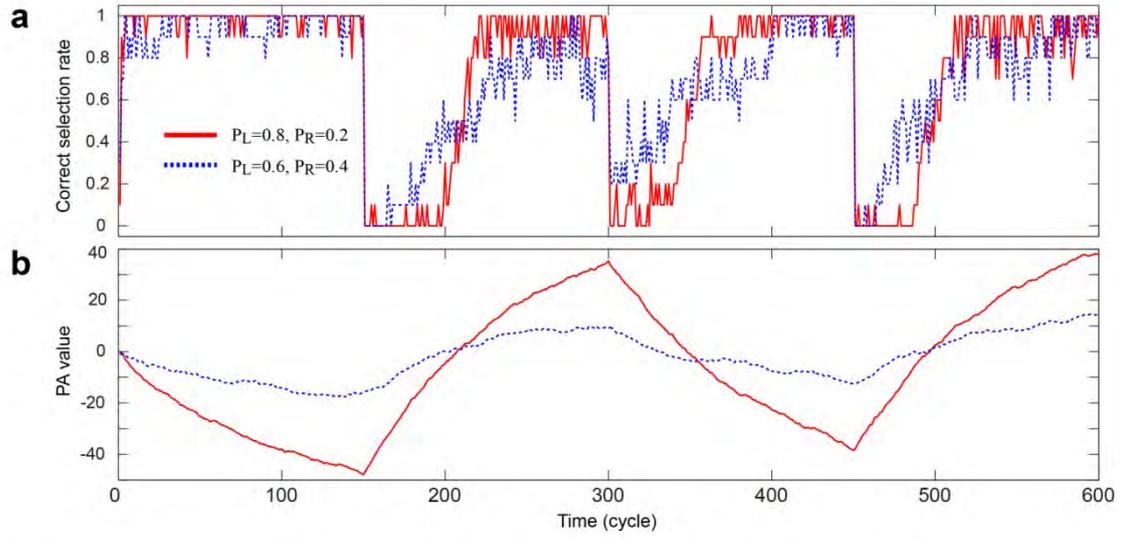

Figure-6 (Naruse)

# Supplementary Information

## Single-photon decision maker

Makoto Naruse[1], Martin Berthel[2,3], Aurélien Drezet[2,3], Serge Huant[2,3], Masashi Aono[4,5], Hirokazu Hori[6] & Song-Ju Kim[7]

1 Photonic Network Research Institute, National Institute of Information and Communications Technology, 4-2-1 Nukui-kita, Koganei, Tokyo 184-8795, Japan

2 Université Grenoble Alpes, Inst. NEEL, F-38000 Grenoble, France

3 CNRS, Inst. NEEL, F-38042 Grenoble, France

4 Earth-Life Science Institute, Tokyo Institute of Technology, 2-12-1 Ookayama, Meguru-ku, Tokyo 152-8550, Japan

5 PRESTO, Japan Science and Technology Agency, 4-1-8 Honcho, Kawaguchi-shi, Saitama 332-0012, Japan

6 Interdisciplinary Graduate School of Medicine and Engineering, University of Yamanashi, Takeda, Kofu, Yamanashi 400-8511, Japan

7 WPI Center for Materials Nanoarchitectonics, National Institute for Materials Science, 1-1 Namiki, Tsukuba, Ibaraki 305-0044, Japan

**Second-order photon-intensity correlation measurements**

The image shown in Supplementary Fig. 1a is a confocal microscope image of the single NV centre used in the main text, and the light blue curve in Supplementary Fig. 1b shows the result of measuring the second-order photon correlation from the single NV centre. The measurement was made using a standard Hanbury–Brown and Twiss correlator. The thick red curve is a fit with a three-level model[1]. The antibunching dip, which is a signature of the quantum nature of the emission, does not drop to zero at a finite delay as expected from a single-photon source because the detectors receive some spurious background fluorescence B ≈ 2000 photons/s in addition to the useful (quantum) emission S ≈ 5000 photons/s from the single NV centre. Taking into account this background in the three-level fit as explained in ref. 1 gives an antibunching dip at zero delay dropping to approximately 0.5, in agreement with the experiment.

**Dependence of single-photon decision maker on strategy**

As discussed in the Methods section, the orientation of the linear polarizer based on the truncated integer values of the PA values are given by, for the **Control 1** policy, $\text{Pos}(-3) = 60$, $\text{Pos}(-2) = 59$, $\text{Pos}(-1) = 58$, $\text{Pos}(0) = 34$, $\text{Pos}(1) = 21$, $\text{Pos}(2) = 20$, and $\text{Pos}(3) = 19$, which we refer to hereafter as **Strategy 1**. Let us define the effective extinction ratio (EER) by the number of the photon counts detected in Ch.0 divided by those detected in Ch.1, which are experimentally measured, as shown in Fig. 4. The correspondence between the truncated PA values and the EER of **Strategy 1** is indicated by the red square marks in the inset of Supplementary Fig. 2, where EER is more sensitive around the origin of the truncated PA value $[\lceil PA(t) \rceil = 0]$. We can think of other two contrasting strategies where (1) EER changes gradually as a function of the truncated PA values (**Strategy 2**) and (2) the change in EER is sensitive at the boundary of the range of the truncated PA values (**Strategy 3**), which are shown by green circles and blue triangles, respectively. Specifically, **Strategies 2** and **3** are, respectively, defined by

**Strategy 2**: $\text{Pos}(-3) = 60$, $\text{Pos}(-2) = 51$, $\text{Pos}(-1) = 42$, $\text{Pos}(0) = 34$, $\text{Pos}(1) = 29$, $\text{Pos}(2) = 24$, and $\text{Pos}(3) = 19$

**Strategy 3**: $\text{Pos}(-3) = 60$, $\text{Pos}(-2) = 36$, $\text{Pos}(-1) = 35$, $\text{Pos}(0) = 34$, $\text{Pos}(1) = 33$, $\text{Pos}(2) = 32$, and $\text{Pos}(3) = 19$.

The resulting decision-making performance of **Strategies 1, 2,** and **3** in terms of correct selection rate is shown by the red solid, green dashed, and blue dotted curves, respectively, in Supplementary Fig. 2, where no evident difference in performance is observed. Based on this evaluation and by considering a strategy in which EER exhibits a dramatic change around *PA* = 0, we choose **Strategy 1** for the experimental demonstration discussed in the main text.

**Supplementary references**

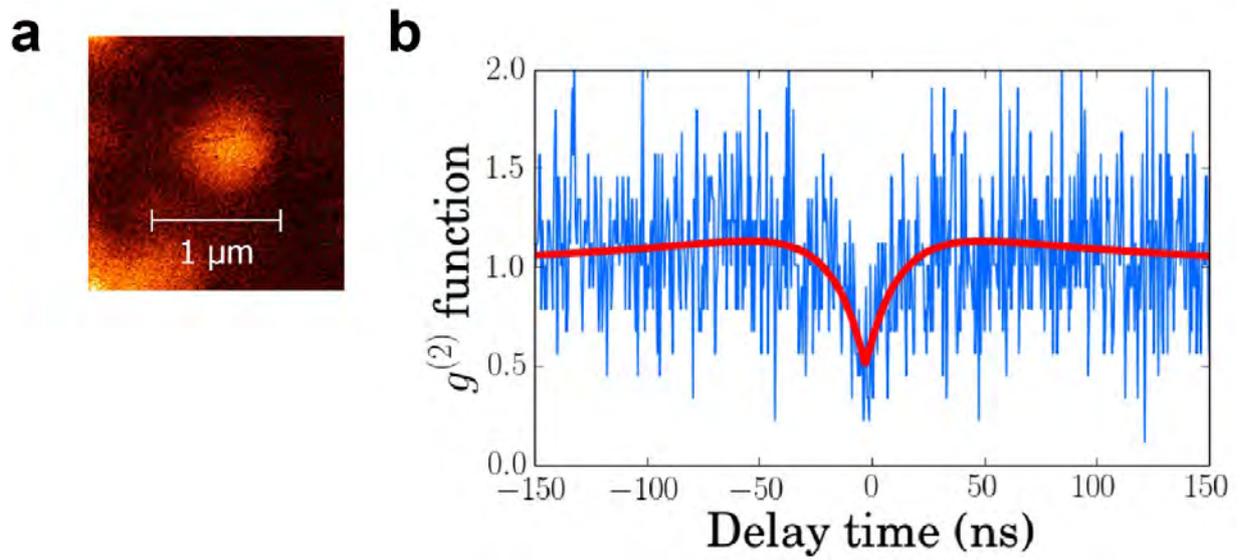

**Supplementary Figure 1 | Second-order photon–intensity correlation measurement of the single NV centre.** (a) A confocal microscope image of the single NV centre used in the main text. (b) The red curve is fit to three-level model.

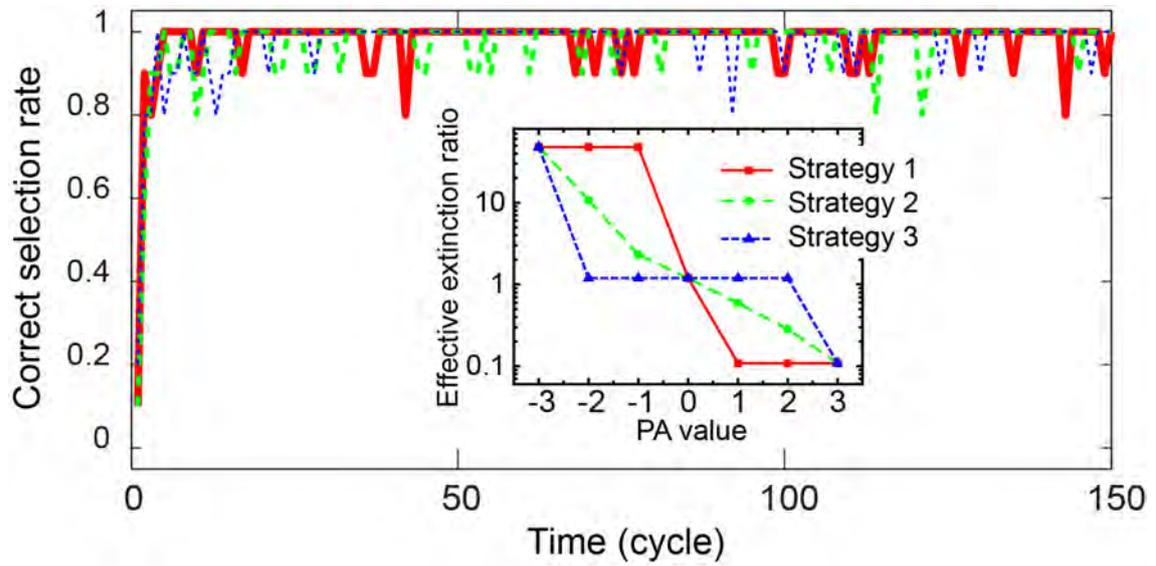

**Supplementary Figure 2 | Dependence of single-photon decision maker on strategy.** The resulting decision-making performance of **Strategies 1, 2**, and **3** in terms of correct selection rate is shown by the red solid, green dashed, and blue dotted curves.